\documentclass[11pt]{article}
\usepackage{moriond,epsfig}
\usepackage{amsmath,amssymb}
\usepackage[all]{xy}
\usepackage{slashed}
\usepackage{psfrag}

\begin{document}
\begin{tabular}{l}
IPPP/09/39\\
DCPT/09/78
\end{tabular}

\vspace*{4cm}
\title{$B\to K^*\mu^+\mu^-$: SM AND BEYOND}

\author{ AOIFE BHARUCHA}

\address{IPPP, Department of Physics,
University of Durham, Durham DH1 3LE, UK\\ \vspace*{.5 cm}
{\it Invited talk at the XLIVth Rencontres de Moriond on Electroweak Interactions\\ and Unified Theories, La Thuile, Italy, 7-14 Mar 2009.}}

\maketitle\abstracts{
The rare decay $B\to K^*\mu^+\mu^-$ is considered to be a golden channel for LHCb, as the polarization of the $K^*$ allows a precise angular reconstruction resulting in many observables that offer new important tests of the Standard Model and its extensions. These angular observables can be expressed in terms of CP-conserving and CP-violating quantities which we study in terms of the full form factors calculated from QCD sum rules on the light-cone, including QCD factorization corrections. We investigate
all observables in the context of the Standard Model and various New Physics models, in
particular the Littlest Higgs model with T-parity and various MSSM scenarios, identifying
those observables with small to moderate dependence on hadronic quantities and large impact
of New Physics.}

\section{Introduction}
In an attempt to identify observables that can shed light on Physics at the TeV scale, a common starting place is the decay $B\to K^*(\to K\pi) \mu^+\mu^-$. This is because it is a flavour changing neutral current(FCNC) process, so the leading order diagrams are at the one-loop level, which raises the chances of New Physics (NP) particles creating observable effects. However, as opposed to the other popular FCNC $b \to s \gamma$, the four body final state gives rise to a multitude of promising angular observables. In addition, this channel satisfies the preference of the LHC for exclusive modes.

This decay has been intensively studied in many papers, we can mention only a few. In '99 Ali et al. calculated the di-lepton mass spectrum and the forward backward asymmetry ($A_{FB}$)
 for the Standard Model (SM) and several Minimal Supersymmetric Model (MSSM) scenarios, using Na\"{i}ve factorization and QCD sum rules on the light cone~\cite{Ali99}. This was extended by Beneke et al.~\cite{BFS01} in the QCD Factorization (QCDF) framework, to include NLO corrections in $\alpha_s$. More recently two further papers appeared, concentrating on observables~\cite{Hiller08,Hurth08}.

The following is based on the recent paper by Altmannshofer et al.~\cite{Altmannshofer:2008dz}. In this work we use the full set of form factors, rather than the two form factors in the heavy
quark limit, calculated from QCD sum rules on the light-cone (LCSR). We study all angular observables in the SM and a variety of NP models and also identify those with small sensitivity to hadronic and large sensitivity to NP effects, with attention to the effects of additional operators, which are extremely suppressed or do not exist in the SM, on all angular observables. In the following we present a collection of interesting results from this work.

\section{Angular observables}\label{sec:AngularObservables}
The differential decay rate for the final state can be expressed \cite{Kruger:1999xa} as
\begin{equation}
\frac{{\rm d}^4\Gamma}{{\rm d}q^2 {\rm d}\cos\theta_l\,{\rm d}\cos\theta_{K}\,{\rm d}\phi)a}=\frac{9}{32\pi}I(q^2,\theta_l,\theta_{K},\phi)
\end{equation}
where 
\begin{eqnarray}
\nonumber I(q^2, \theta_l, \theta_K, \phi)&=&I_1^s \sin^2\theta_K + I_1^c \cos^2\theta_K+ (I_2^s \sin^2\theta_K + I_2^c \cos^2\theta_K) \cos 2\theta_l\\
\nonumber&&+ I_3 \sin^2\theta_K \sin^2\theta_l \cos 2\phi+I_4 \sin 2\theta_K \sin 2\theta_l \cos\phi+ I_5 \sin 2\theta_K \sin\theta_l \cos\phi\\
&&+(I_6^s \sin^2\theta_K+I_6^c \cos^2\theta_K )\cos\theta_l \nonumber+ (I_7 \sin\theta_l+ I_8\sin 2\theta_l) \sin 2\theta_K \sin\phi\\
&&+ I_9 \sin^2\theta_K \sin^2\theta_l \sin 2\phi
\label{eq:Is}
\end{eqnarray}
Here $q^2$ is the invariant mass of the muons. In order to define the angles it is simplest to first define the z axis by the direction of the $K^*$ in the B meson rest frame. Then $\theta_l$ is the angle made by the $\mu^-$ with respect to the negative direction of the z axis in the centre of mass frame of the muons, $\theta_K$ is the direction of the K meson with respect to the z axis in the centre of mass frame of the $K^*$ and $\phi$ is the angle between the planes defined by the centre of mass of the muons and the $K$ and $\pi$. In analogy we can define $\bar{I}^{(a)}_{i}$ by taking the CP conjugate of Eq.~(\ref{eq:Is}), and making the replacements
\begin{equation}
I^{(a)}_{1,2,3,4,7}\to\bar{I}^{(a)}_{1,2,3,4,7} \qquad{\rm and}\qquad I^{(a)}_{5,6,8,9}\to-\bar{I}^{(a)}_{5,6,8,9}.
\end{equation}
The angular coefficients $\bar{I}^{(a)}_{i}$($I^{(a)}_{i}$), are all observable quantities. However, we choose instead to study two different sets of observables. In order to emphasize CP conserving effects and CP violating effects \footnote{Our definition for the $A_i^{(a)}$'s differs from that of Bobeth et al.~\cite{Hiller08} by a factor $3/2$} we define~\cite{Altmannshofer:2008dz}
\begin{equation}
S_i^{(a)}=\frac{I_i^{(a)}+\bar{I}_i^{(a)}}{d(\Gamma+\bar{\Gamma})/dq^2}\qquad{\rm and}\qquad A_i^{(a)}=\frac{I_i^{(a)}-\bar{I}_i^{(a)}}{d(\Gamma+\bar{\Gamma})/dq^2}
\label{eq:Ss}
\end{equation}
These sets of observables can be extracted from a full angular fit to the decay distribution of appropriate combinations of $B\to K^* \mu^+\mu^-$ and $\bar{B}\to K^* \mu^+\mu^-$, providing information about $q^2$ and the angles $\theta_l$, $\theta_{K^*}$ and $\phi$ is known. Alternatively they can be extracted individually through taking appropriate asymmetries over an angle or combination of angles. This has previously been described in detail with explicit formulae~\cite{Altmannshofer:2008dz}. 

\section{Calculation of angular observables}
\subsection{Wilson coefficients}
The angular observables defined above have been calculated in terms of form factors and Wilson coefficients~\cite{Altmannshofer:2008dz}, including next-to-leading-order corrections given by QCD factorization. The Wilson coefficients are defined using the effective Hamiltonian for $b\to s l^+ l^-$ transitions given by~\cite{bobeth}
\begin{equation} \label{eq:Heff}
    {\cal H}_{\rm{eff}} = - \frac{4\,G_F}{\sqrt{2}}\left(
\lambda_t {\cal H}_{\rm{eff}}^{(t)} + \lambda_u {\cal
  H}_{\rm{eff}}^{(u)}\right)
\end{equation}
where $\lambda_i=V_{ib}V_{is}^*$ and
\begin{eqnarray*}
{\cal H}_{\rm{eff}}^{(t)} 
& = & 
C_1 \mathcal O_1^c + C_2 \mathcal O_2^c + \sum_{i=3}^{6} C_i 
\mathcal O_i + \sum_{i=7-10,P,S} (C_i \mathcal O_i + C'_i \mathcal
O'_i)\,,
\\
{\cal H}_{\rm{eff}}^{(u)} 
& = & 
C_1 (\mathcal O_1^c-\mathcal O_1^u)  + C_2(\mathcal O_2^c-\mathcal
O_2^u)\,.
\end{eqnarray*}
${\cal H}_{\rm{eff}}^{(u)}$ is often neglected, as $\lambda_u$ is doubly Cabbibo suppressed. However $\lambda_u$ has a weak phase, and therefore this amplitude would be important if a NP model with additional CP phases produced effects of a comparable size.
The Wilson coefficients are calculated at next-to-next-to-leading logarithmic (NNLL)
accuracy, involving a calculation of the matching conditions at $\mu=m_W$ to two-loop accuracy~\cite{bobeth}. NP contributions, on the other hand, will be included
to one-loop accuracy only, as calculations show that they are small in the MSSM~\cite{bobeth}. These coefficients must then be rescaled from the electroweak scale to the required physical scale $m_b$, which is achieved using the anomalous dimension matrices which have been calculated to three-loop accuracy~\cite{gorbahn}. The numerical values of the Wilson coefficient used can be found in the paper by Altmannshofer et al.

\subsection{Form factors}\label{sec:formfactors}
$B\to K^* \mu^+\mu^-$ is characterised by a set of eight form factors. These are hadronic quantities, and for certain ranges in $q^2$ can be obtained by non-perturbative calculation. QCD sum rules on the light cone (LCSR) is a well established alternative technique which provides the only set of results for the desired range in  $q^2$ ~\cite{BZ04}. The method involves combining classic QCD sum rules \cite{SVZ} with information on light cone distribution amplitudes. In the large energy limit of the $K^*$ these form factors satisfy certain relations, and reduce to two soft form factors $\xi_\perp$ and $\xi_\parallel$ used in the QCDF framework~\cite{BFS01}. The relations were studied through appropriate ratios of the LCSR predictions for the full form factors~\cite{Altmannshofer:2008dz}, and found those involving $\xi_\perp$ are almost independent of $q^2$, but those involving  $\xi_\parallel$ had a definite dependence on $q^2$, so are probably more sensitive to the $1/m_b$ corrections neglected in QCDF.

\subsection{QCD factorization}
The angular coefficients are functions of transversity amplitudes, which can be expressed in terms of the full form factors~\cite{Kruger:2005ep,Hiller08}. In addition to these expression we include the QCDF corrections, NLO in $\alpha_s$ but LO  in $1/m_b$~\cite{Altmannshofer:2008dz}. These corrections correspond to the weak annihilation and non-factorizable contributions to the transversity amplitudes, in accordance with previous notation\cite{BFS01,LunghiMatias}. There are two types of $O(\alpha_s)$ corrections, factorizable and non-factorizable.  Both are normally included, but we only include the non-factorizable corrections. The factorizable corrections arise on expressing the full form factors in terms of $\xi_\perp$ and $\xi_\parallel$. We express our leading order results in terms of the full form factors, automatically including these factorizable corrections as well as the most important source of $O(1/m_b)$ corrections, as argued in \cite{Altmannshofer:2008dz}. The weak annihilation correction is induced by the penguin operators $O_3$ and $O_4$ and hence is numerically small. It is leading in $1/m_b$ and $O(\alpha_s)$, so in principle one should also include power-suppressed and radiative corrections. However, in view of the small size of such corrections, we feel justified in neglecting them.

\section{New Physics contribution}
We study effects of specific NP scenarios. We have chosen models which differ in terms of additional operators, CP violation beyond the SM, and non-standard flavour structure. We try to identify observables which might have particular sensitivity to these properties, and explore whether such effects might be observable. 

\subsection{Overview of models}
We have included four sets of Wilson coefficients in a variety of NP scenarios. These are intended to give a feeling for the possible effects of NP on the numerous observables. The models chosen provide a wide range of possibilities of the effects of NP on the observables. We describe here the models only in terms of the properties listed above, and the theoretical motivation and details of these models is found in the references mentioned below.
\begin{itemize}
\item \textbf{Minimal Flavour Violation}(MFV): We first consider constrained minimal flavour violation (CMFV)~\cite{Buras:2003jf}. Here there are no additional operators or phases, and the CP and flavour structure of the SM is preserved, so CP and flavour violating observables are generally SM-like. We also consider the minimal flavour violating MSSM (MFVMSSM)\cite{LMSS99}, where scalar and pseudoscalar operators arise due to the non-standard Higgs structure. 
\item \textbf{Flavour Blind MSSM}(FBMSSM): Here the minimal flavour violating MSSM is modified by some flavour conserving but CP violating phases in the soft SUSY breaking trilinear couplings~\cite{ABP08}. This results in complex phases in the Wilson coefficients, in particular $C_7$ is affected. 
\item \textbf{General MSSM}(GMSSM): Minimal flavour violation is not imposed, and generic flavour- and CP-violating soft SUSY-breaking terms are allowed~\cite{Gabrielli:2002me}. The parameters are only constrained by experimental bounds, and interestingly, these bounds allow a large complex $C_7^\prime$. We emphasize how this has an impact on the observables in the phenomenological analysis.
\item\textbf{Littlest Higgs Model with T-parity}(LHT): FCNC interactions involving SM quarks, heavy mirror quarks and new heavy charged and neutral gauge bosons occur via a new mixing
matrix with three new mixing angles and three new CP-violating phases~\cite{Cheng:2004yc,LHT2}. This leads to the possibility of a large complex $C_9$ and $C_{10}$.
\end{itemize}

\section{Results and Conclusions}\label{subsec:results}
In our previous work~\cite{Altmannshofer:2008dz} we show our SM predictions for the observables $S^{(a)}_i$ and $A^{(a)}_i$ respectively. In the SM $S_{1/2}^{s/c}$ are numerically the largest quantities, and constitute the bulk of the di-lepton mass distribution. $S_{4}$, $S_{5}$, $S_{6}^s$ are also sizeable and particularly interesting, as they all have a zero in $q^2$. $S_{6}^s$, up to a normalization factor, is the CP averaged forward backward asymmetry $A_{FB}$, however $S_{4}$ and $S_{5}$ have only previously been studied in the context of  $A_T^{(3)}$ and $A_T^{(4)}$ defined previously~\cite{Hurth08} , where they are combined with other $S_i$'s. $S_3$ is very small in the SM as it is sensitive to the Wilson coefficients $C'_7$ which is suppressed by a factor of $m_s/m_b$ in the SM. All the above observables are protected from hadronic effects by the normalization to the CP averaged total decay rate. As predicted, the CP asymmetries are close to vanishing in the SM~\cite{Hiller08}.

\begin{figure}[htb]
\centering
\psfig{figure=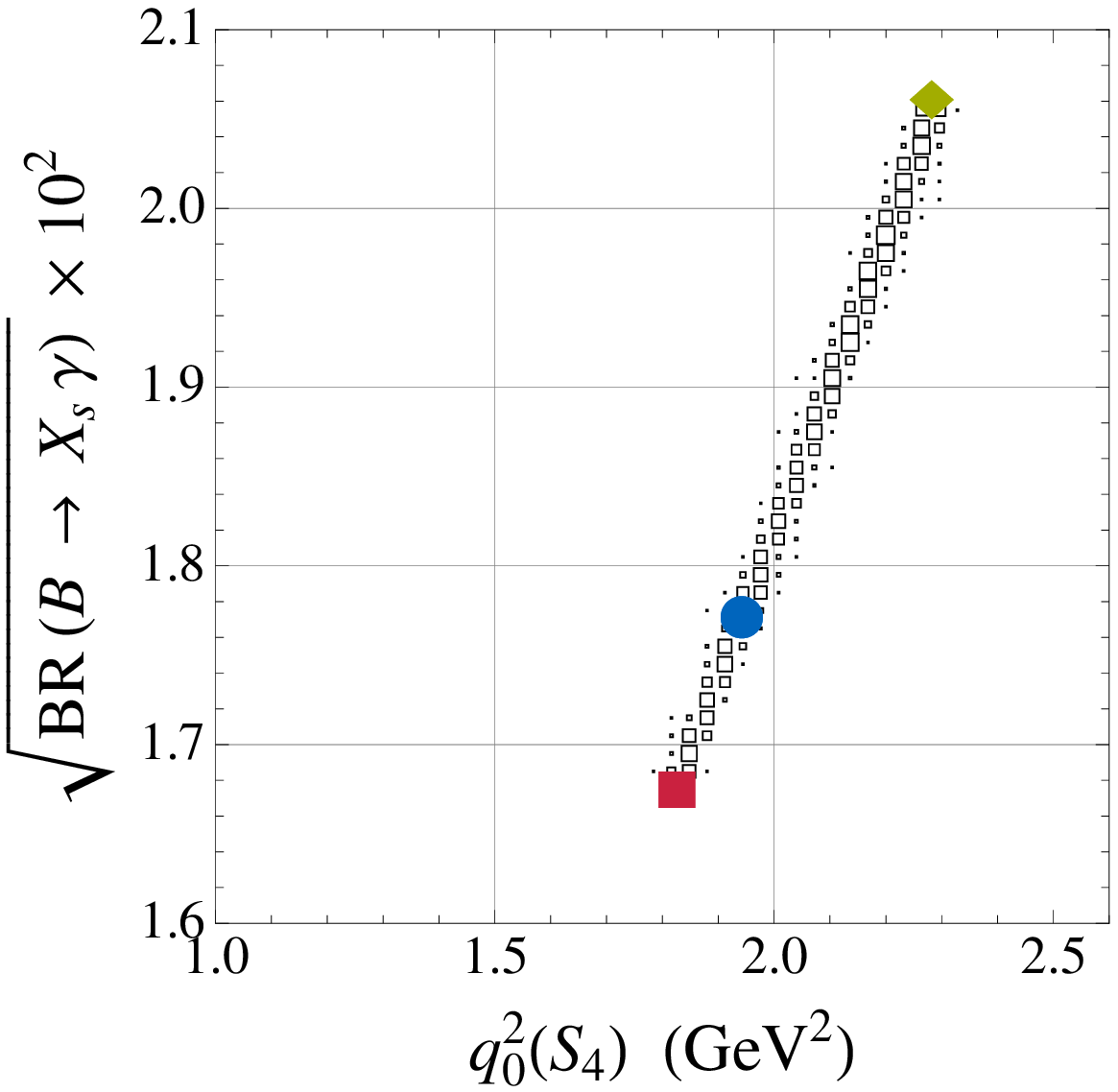,height=1.5in}
\psfig{figure=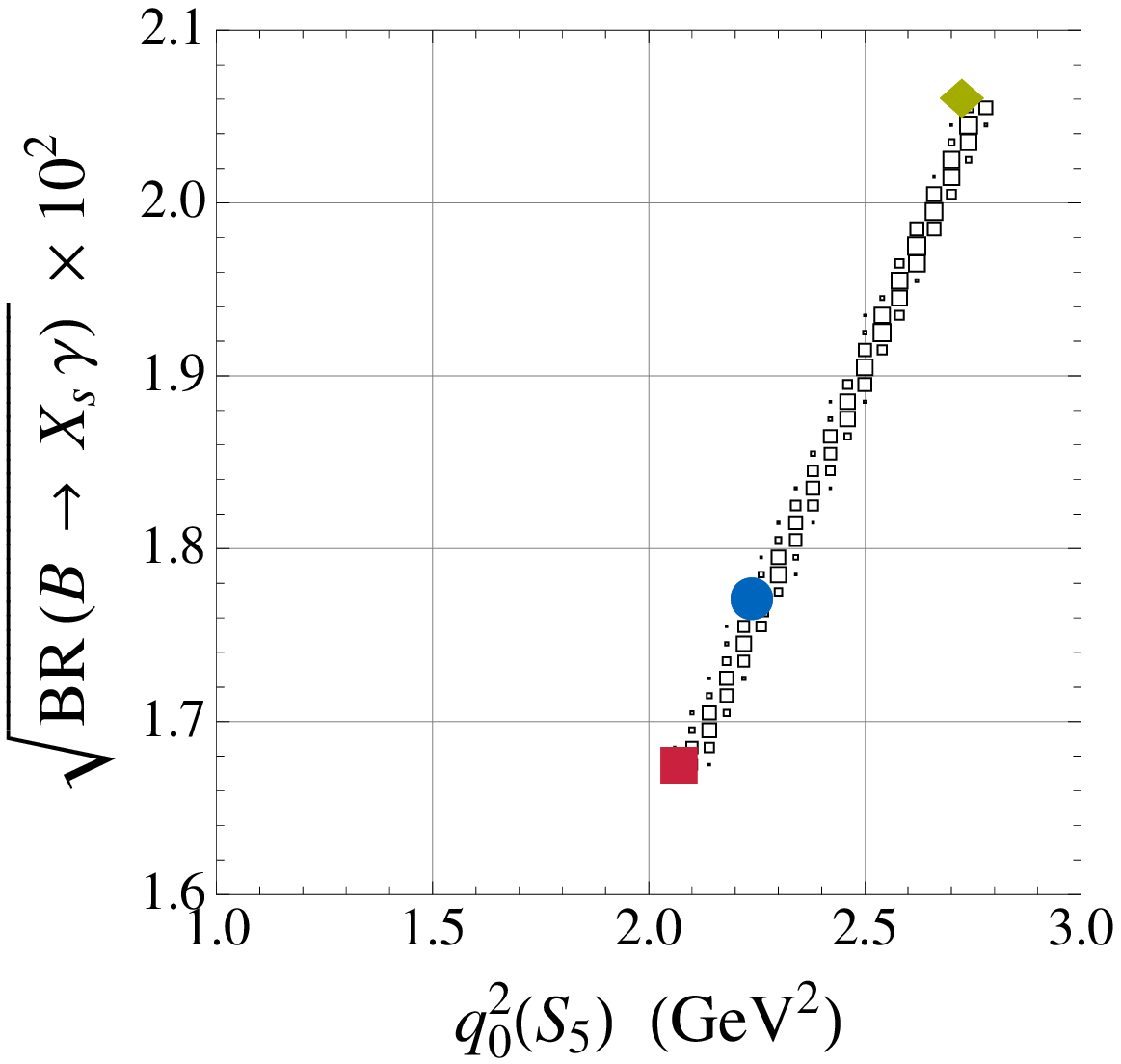,height=1.5in}
\psfig{figure=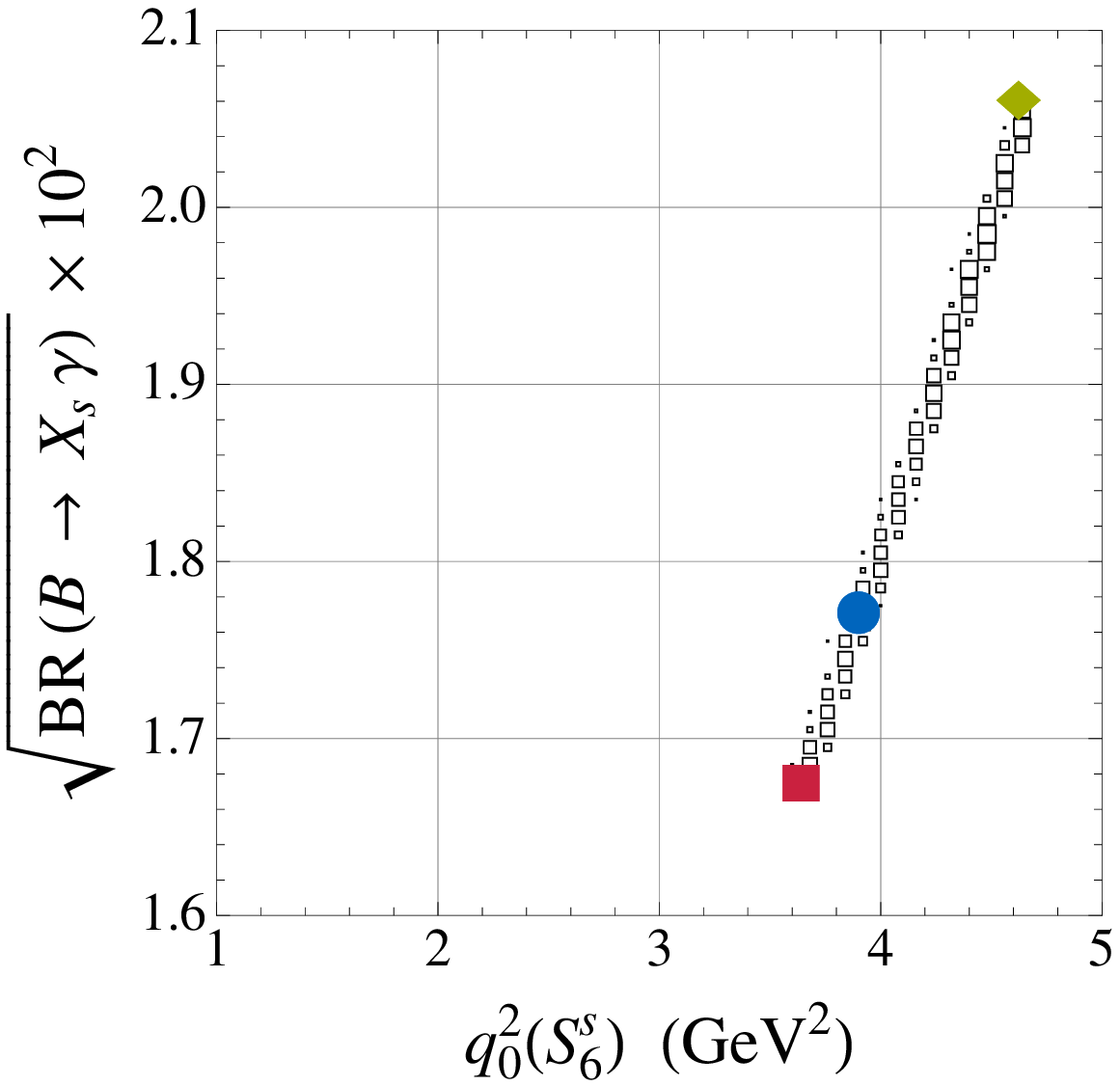,height=1.5in}
\caption[]{\small The correlation between the zeros of $S_4$, $S_5$
  and $S_6^s$ and BR$(B \to X_s \gamma)$ in the MFV MSSM. The blue
  circles correspond to the central SM values, while the green
  diamonds represent scenario MFV$_{\rm I}$ and the red squares
  scenario MFV$_{\rm II}$\cite{Altmannshofer:2008dz}.}
\label{fig:CMFV}
\end{figure}
As stated earlier in CMFV there are no additional operators or any non-SM CP or flavour violating structure. Effects for CMFV are found to be in general small, though in some cases can reach $50\%$. A correlation plot of the position of the zeros of $S_4$, $S_5$, $S_6^s$ with $\sqrt{B(b\to s\gamma)}$ is seen in Fig.~\ref{fig:CMFV}. 
\begin{figure}[htb]
\centering
\psfig{figure=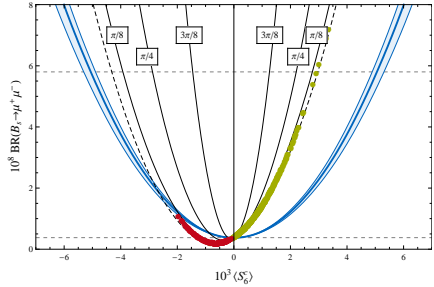,height=1.5in}
\label{fig:MFVMSSM}
\caption[]{Correlation between the observable $S_6^c$ and the branching ratio of $B_s\to\mu^+\mu^-$. The blue (dark grey) band is obtained by assuming NP contributions only to the Wilson coefficient $C_S$, the light grey band by assuming $C_P=-C_S$. The red (dark grey) dots correspond to points in the CMSSM as described in the text. The horizontal dashed lines indicate the SM prediction and the current experimental upper bound for $\rm{BR}(B_s\to\mu^+\mu^-)$ \cite{Altmannshofer:2008dz}.}
\end{figure}
In the MFVMSSM, the Scalar operators can have interesting effects, as the observable $S_6^c$ is only non-zero in the presence of such operators. The current bounds on these operators come from the branching ratio of $B_s\to\mu^+\mu^-$, as seen in Fig.~\ref{fig:MFVMSSM} we show the correlation between $S_6^c$ and $\rm{BR}(B_s\to\mu^+\mu^-)$.

\begin{figure}[tb]
\centering
\psfig{figure=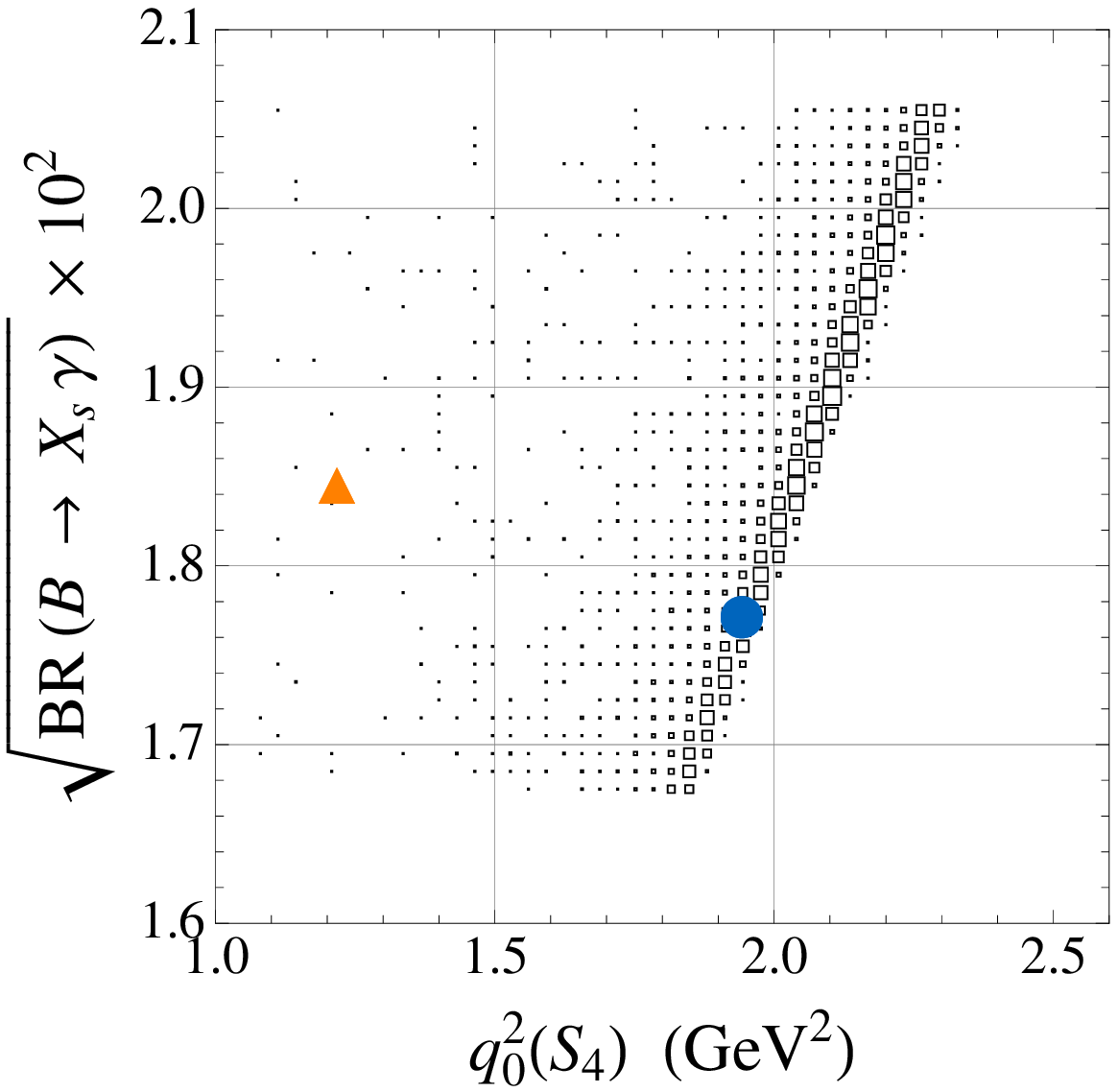,height=1.5in}
\psfig{figure=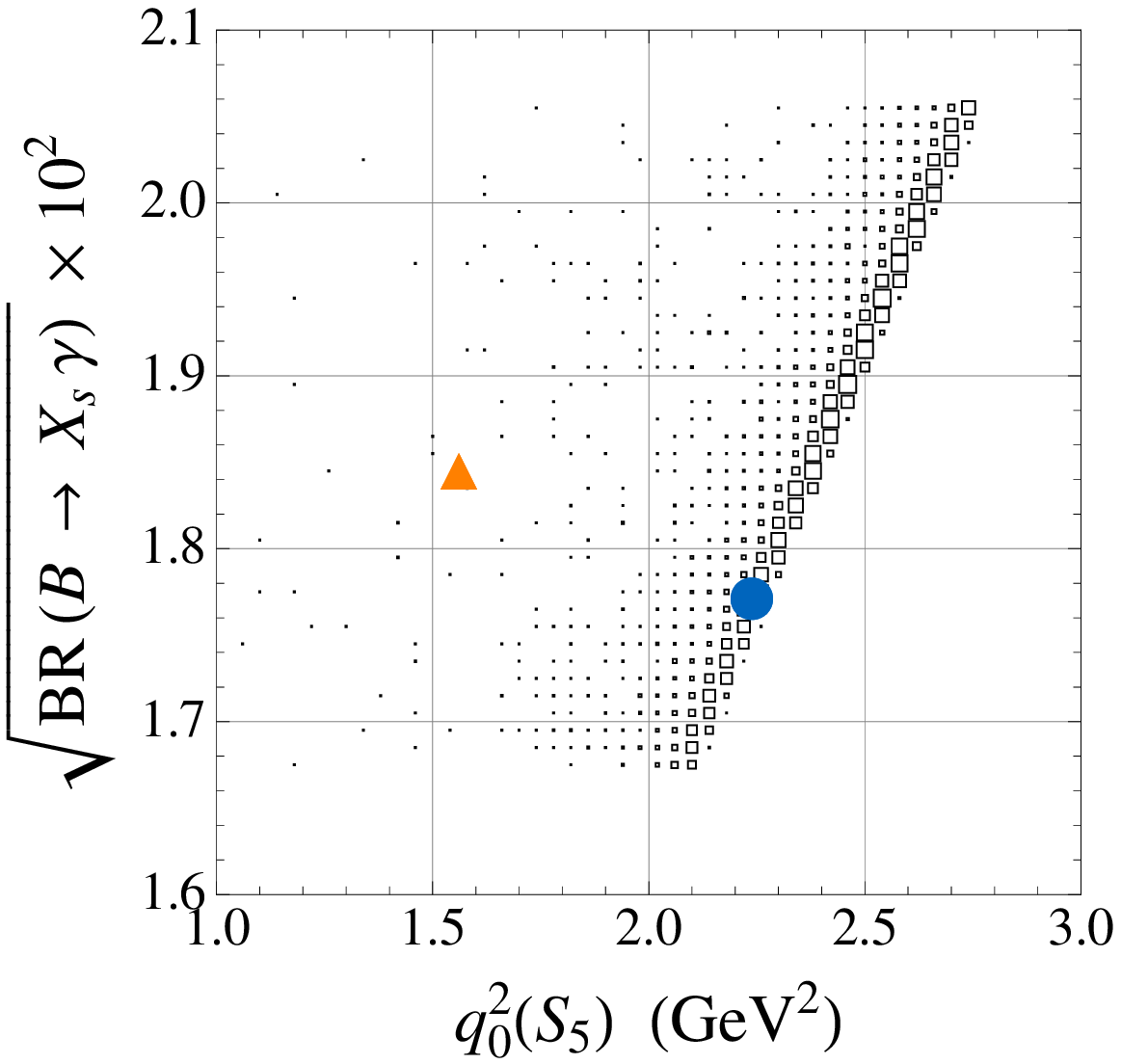,height=1.5in}
\psfig{figure=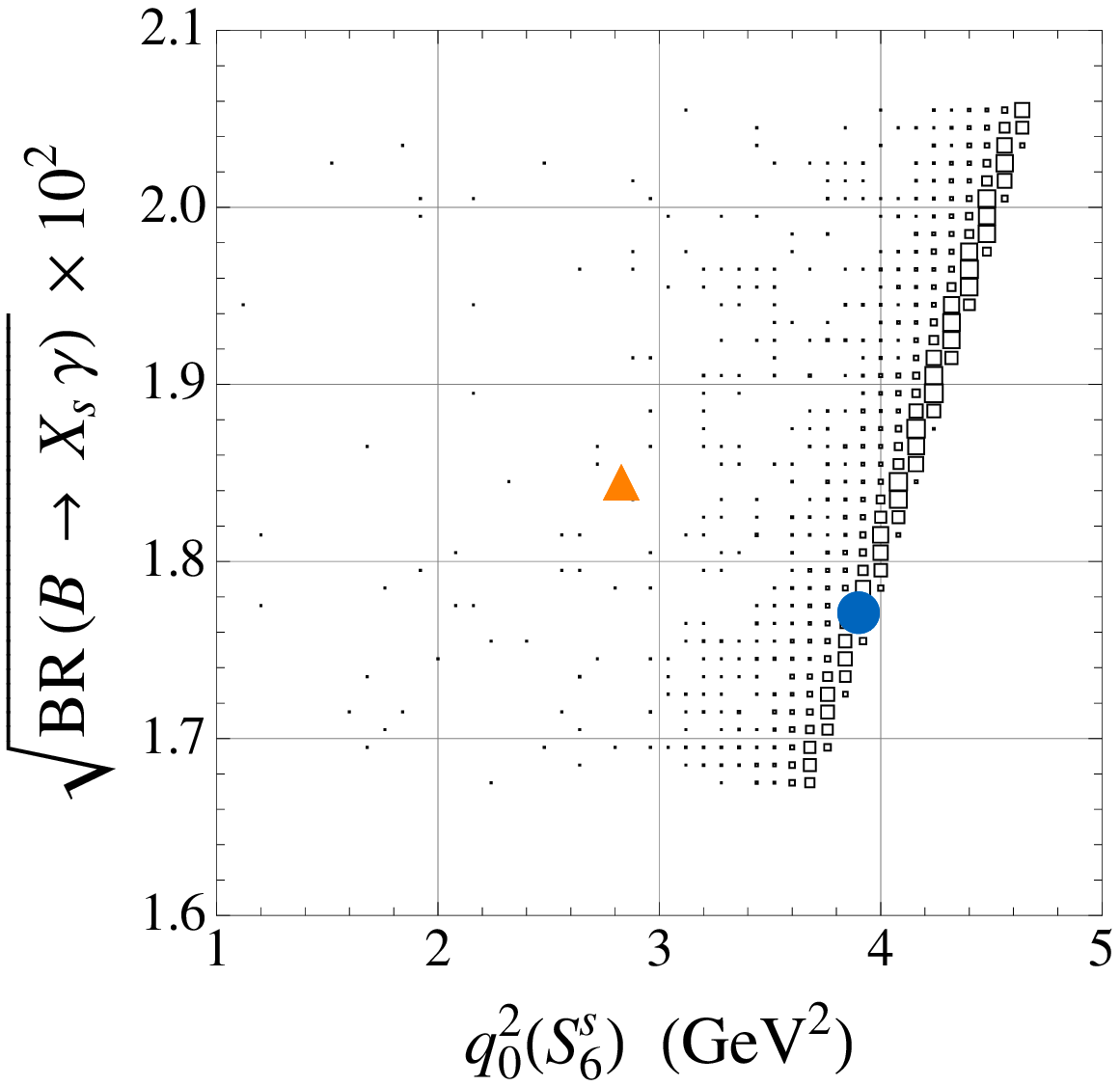,height=1.5in}
\caption[]{Correlation between the zeros of $S_4$, $S_5$ and $S_6^s$ with the $b \to s \gamma$ branching ratio (upper plots) and with the integrated asymmetry $\langle A_7 \rangle$ (lower plots) in the FBMSSM. The blue circles correspond to the SM predictions. The orange triangles correspond to a FBMSSM scenario that gives $S_{\phi K_S}$ close to the central experimental value $\simeq 0.44$.}\label{fig:FBMSSM1}
\end{figure}

In the FBMSSM, the Wilson coefficient $C_7$ may be complex due to the additional CP violating phases introduced in the trilinear couplings. This has interesting consequences, as the bound on $C_7$ from $b\to s\gamma$ weakened, as there is a cancellation between real and imaginary contributions. Therefore when we plot the correlation between the position of the zeros of $S_4$, $S_5$, $S_6^s$ with $\sqrt{B(b\to s\gamma)}$ in Fig.~\ref{fig:FBMSSM1}, there is seen to be more freedom than in Fig.~\ref{fig:CMFV}, and in some cases the zero can even vanish.

\begin{figure}[htb]
\centering
\psfig{figure=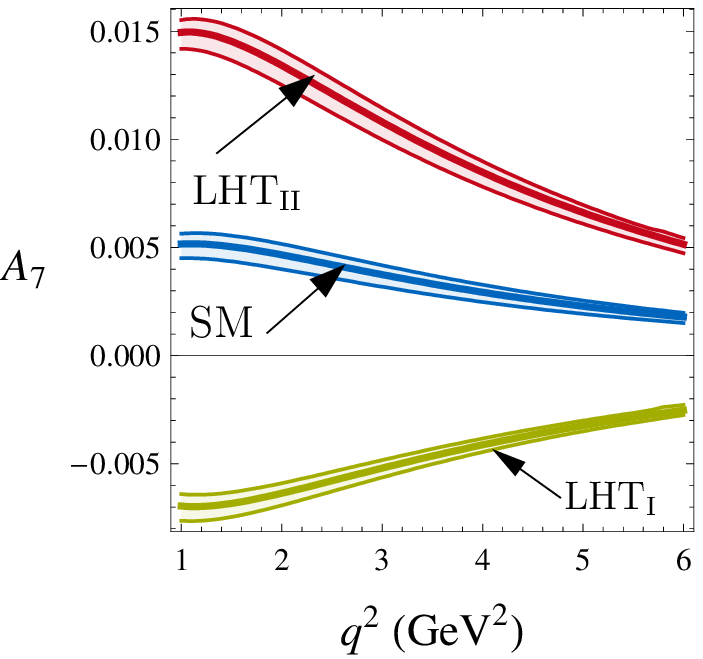,height=1.5in}\psfig{figure=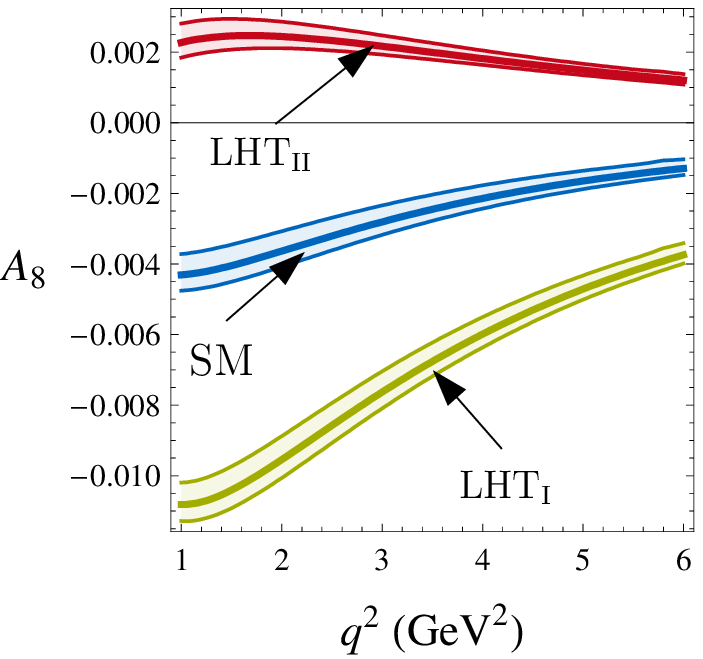,height=1.5in}
\caption{CP asymmetries $A_7$ and $A_8$ in the SM (blue band) and the LHT
scenarios LHTI,II.
\label{fig:FBMSSM2}}
\end{figure}
\begin{figure}
\centering
\psfig{figure=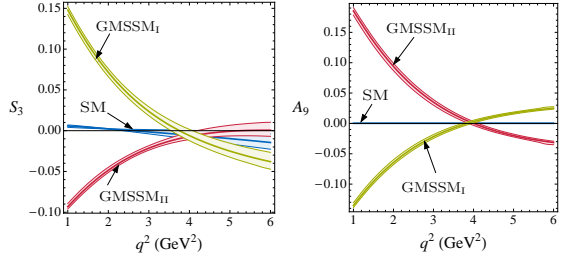,height=1.5in}
\caption{The observables $S_3$ and $A_9$ in the SM (blue band) and the two GMSSM scenarios
GMSSMI,II with large complex contributions to $C_7^\prime$ as described in the text.
\label{fig:GMSSM}}
\end{figure}
The GMSSM differs from other MSSM models considered, as a large complex $C_7^\prime$ can be generated via down squark gluino loops. We choose to concentrate on how this affects the observables, and find sizeable effects in $S^{(i)}_{4/5/6}$, $A_{7/8}$, and uniquely in $S_3$/$A_9$, seen in Fig.~\ref{fig:GMSSM}.
We conclude that $B\to K^*\mu^+\mu^-$ will play a key role in the search for beyond the Standard Model effects. In particular it can shed light on any additional operators or CP violation, and on non-standard flavour structure. Studies are currently ongoing at LHCb to determine experimental sensitivity to these observables~\cite{Gibson}.

\section*{Acknowledgments}
A.K.M.B. is grateful to the organisers of the conference, to her supervisor Patricia Ball and her collaborators Wolfgang Altmannshofer, Andrzej Buras, David Straub and Michael Wick. This work was supported by a UK STFC studentship, additional financial support from Lehrstuhl T31, TU Munich and the EU network contract No. MRTN-CT-2006-035482(Flavianet).

\end{document}